\documentclass[aps, prb, showpacs, twocolumn, amsfonts, amsmath, amssymb, superscriptaddress, floatfix]{revtex4-1}

\usepackage[T1]{fontenc}
\usepackage[latin9]{inputenc}
\usepackage{color}
\usepackage{graphicx}
\usepackage{dsfont}

\def\be{\begin{equation}}
\def\ee{\end{equation}}
\def\bea{\begin{eqnarray}}
\def\eea{\end{eqnarray}}

\begin{document}

\title{Mobility edge of two interacting particles in three-dimensional random potentials}

\author{Filippo Stellin} 
%\affiliation{Laboratoire Mat\' eriaux et Ph\'enom\`enes Quantiques, Universit\' e Paris Diderot-Paris 7 and CNRS, UMR 7162, 75205 Paris Cedex 13, France}
\affiliation{Universit\' e de Paris, Laboratoire Mat\' eriaux et Ph\' enom\`enes Quantiques, CNRS, F-75013, Paris, France}
\author{Giuliano Orso}
\affiliation{Universit\' e de Paris, Laboratoire Mat\' eriaux et Ph\' enom\`enes Quantiques, CNRS, F-75013, Paris, France}

%\pacs{03.75.-b, 05.30.Rt, 64.70.Tg, 05.60.Gg}

%\date{Version 1 dated 15 April, 2016}

\begin{abstract}

We investigate Anderson transitions for a system of two particles moving in a three-dimensional disordered lattice and subject to on-site (Hubbard) interactions of 
strength $U$. The two-body problem is exactly mapped  into an effective single-particle equation for the center of mass motion, whose 
localization properties are studied numerically.    We show that, for zero total energy of the pair, %and uniform distribution of the disorder, 
the  transition occurs in a regime where all single-particle states are localized. In particular the critical disorder strength exhibits a non-monotonic behavior as a function of $|U|$, increasing sharply for 
weak interactions and converging to a finite value in the strong coupling limit. Within our numerical accuracy, 
short-range interactions do not affect the universality class of the  transition.
%Hence the interaction-induced shift of the mobility edge due to disorder screening does not require a finite density of (fermionic) particles.
\end{abstract}

\maketitle

\section{Introduction and motivations}
 Wave diffusion in disordered media can be completely inhibited~\cite{Anderson:LocAnderson:PR58} due to interference effects between the multiple scatterings
 from the randomly distributed impurities.  This phenomenon, known as Anderson  localization, has been observed for several kinds of wave-like systems, 
including light waves in diffusive media~\cite{Wiersma:LightLoc:N97,Maret:AndersonTransLight:PRL06} or
photonic crystals~\cite{Segev:LocAnderson2DLight:N07,Lahini:AndersonLocNonlinPhotonicLattices:PRL08}, 
ultrasound~\cite{vanTiggelen:AndersonSound:NP08}, microwaves~\cite{Chabanov:StatisticalSignaturesPhotonLoc:N00} and 
atomic matter waves~\cite{Billy:AndersonBEC1D:N08,Roati:AubryAndreBEC1D:N08}.

In quantum systems, this effect appears through the spatial localization of  the  wave-functions. 
In the absence of magnetic fields and of spin-orbit couplings, all states are exponentially localized in one and in two dimensions, whereas in three dimensions there exists
a critical value $E_c$ of the particle energy, called mobility edge,  separating localized from extended states. At this point the system undergoes a metal-insulator transition~\cite{Evers:AndersonTransitions:RMP08}. 
Mobility edges have been reported~\cite{Kondov:ThreeDimensionalAnderson:S11,Jendrzejewski:AndersonLoc3D:NP12,Semeghini:2014}  in experiments with 
non-interacting ultracold atoms in three-dimensional (3D) speckle potentials,
and their measured values have been compared against precise numerical 
estimates~\cite{Delande:MobEdgeSpeckle:PRL2014,Pilati:LevelStats:2015, Pasek:3DAndersonSpeckle:PRA2015,Pilati:3DAndersonSpeckle:2015,Pasek:PRL2017}.  Interestingly, Anderson transitions  have also been observed~\cite{Chabe:Anderson:PRL08}  in momentum space, using cold atoms implementations of the quasi-periodic quantum 
kicked rotor,  
%which is equivalent to a 3D disordered system, 
allowing for the first experimental test of universality~\cite{Lopez:ExperimentalTestOfUniversality:PRL12}. 
For a correlated disorder, mobility edges occur even in lower dimensions, as recently  observed~\cite{Luschen:PRL2018}  for atoms in one-dimensional quasi-periodic optical lattices, 
 in agreement with earlier theoretical predictions~\cite{Boers:PRA2007,Li:PRB2017}.

While single-particle Anderson localization is relatively well understood, its generalization to interacting systems, 
called many-body localization, is more recent~\cite{Altshuler:MetalInsulator:ANP06} and is  currently 
the object of intense theoretical and experimental activities~\cite{Nandkishore2015,Abanin:RMP2019,Parameswaran:RPP2018}. 
Perhaps surprisingly,  even the problem of two interacting particles in a random potential is still open.
In  a seminal work~\cite{Shepelyansky:AndLocTIP1D:PRL94}, Shepelyansky showed that, in the presence of a weak  (attractive or repulsive) interaction, a pair can propagate over a distance much larger than the single-particle localization length. 
It was later argued~\cite{Borgonovi:NonLinearity1995,Imry:CohPropTIP:EPL95} that all two-particle states  remain localized in one and two dimensions 
(although with a possibly large localization length), whereas in three dimensions  an Anderson transition to a diffusive phase 
could occur even when all single-particle states are localized.
While several numerical studies~\cite{Weinmann:PRL1995,vonOppen:AndLocTIPDeloc:PRL96,Frahm1999,Roemer:PhysicaE2001,Dias:PhysicaA2014,Lee:PRA2014,Krimer:InterConnDisord2PStates:PRB15,Frahm:EigStructAL1DTIP16}
have confirmed the claim for one-dimensional systems,  the situation is much less clear in higher dimensions, where the computational cost
 limits the system sizes that can be explored. In particular an Anderson transition
was predicted~\cite{Ortugno:AndLocTIPDeloc:EPL99,Roemer1999} to occur in two dimensions
(see also~\cite{Chattaraj2018} for a recent study of the two-particle dynamics in a similar model).

%In this Letter we investigate Anderson transitions in a system of two particles \toadd{- either bosons or fermions in different spin states -} moving in a 3D disordered lattice  and 
% coupled by on-site interactions, as described by the Anderson-Hubbard model. 
% \toadd{Importantly, we focus on pair states with total energy equal to zero, well above the ground state.}   
%Based on large-scale numerical calculations of the transmission amplitude, we compute  
%the precise phase boundary between localized and extended states  in the interaction-disorder plane, 
%\sout{for zero total energy of the pair}. In particular, we find that the two-particle \sout{system} \toadd{transition} is still described by the orthogonal universality class. 

In this work we investigate Anderson transitions in a system of two particles moving in a 3D disordered lattice and 
coupled by on-site interactions.  The particles can be either  bosons
or fermions with different spins in the singlet state.  %We assume that the pair has zero total energy, corresponding to a highly excited state.
Based on large-scale numerical calculations of the transmission amplitude, we compute  
the precise phase boundary between localized and extended states  in the interaction-disorder plane, for a pair with zero total energy (well above the ground state). 
Importantly, we find that the two-particle Anderson transition is still described by the orthogonal universality class. 

In Sec. II we map exactly the two-particle Hamiltonian into an effective single-particle model, Eq.(\ref{integral}), and compute the associated matrix $K$. In Sec. III we explain how to extract the reduced localization length of a pair with zero total energy from transmission amplitude calculations performed in short bars. We then identify the critical point of the Anderson transition via an accurate finite-size scaling analysis. In Sec. IV we present the phase diagram for Anderson localization of the pair in the interaction-disorder plane. 

\section{Effective single-particle model} 
The two-body Hamiltonian can be written as $\hat H=\hat H_0 + \hat U$, where
$\hat U=U\sum_{\mathbf m}|{\mathbf m},{\mathbf m}\rangle \langle {\mathbf m},{\mathbf m}|$ refers to the on-site (Hubbard) interaction  of strength $U$ 
and $\hat H_0$ is the non interacting part. 
The latter can be written as $\hat H_0=\hat H^\textrm{sp} \otimes  \hat{\mathds{1}} +\hat{\mathds{1}}  \otimes \hat H^\textrm{sp}$, where
\begin{equation}\label{Anderson3D}
\hat H^\textrm{sp}= -J \sum_{\mathbf n, i} |\mathbf n+ \mathbf e_i \rangle  \langle \mathbf n| + \sum_{\mathbf n}V_\mathbf n |\mathbf n\rangle \langle \mathbf n| 
\end{equation}
is the single-particle Anderson model.  Here $J$ is the tunneling rate between neighboring sites, 
$\mathbf e_i$  are the unit vectors along the three orthogonal axes  and $V_{\mathbf n}$ is the value of the random potential at site 
$\mathbf n$.  
In the following we fix the energy scale by setting $J=1$  and assume that the random potential is uniformly distributed 
in the interval $[-W/2,W/2]$. Then all single-particle states are localized for
$W>W_c^{sp}=16.54 \pm 0.01$~\cite{McKinnonKramer:TransferMatrix:ZPB83,Slevin:CriticalExponent:NJP14}.

The Schr\" odinger equation for the pair can be written as $(E -\hat H_0)|\psi\rangle=\hat U|\psi\rangle$, $E$ being the total energy. 
 Applying the Green's function operator $\hat G(E)=(E \hat I -\hat H_0)^{-1}$
 to both sides of this equation, we find   
 \begin{equation}
\label{formalism2}
|\psi\rangle=\sum_{\mathbf m} U \hat G(E) |{\mathbf m},{\mathbf m}\rangle \langle {\mathbf m},{\mathbf m}|\psi\rangle,  
\end{equation}
showing that the  wave-function can be completely reconstructed from the diagonal amplitudes
$f_{\mathbf m}=\langle {\mathbf m},{\mathbf m}|\psi\rangle$. By projecting Eq.(\ref{formalism2}) over the state 
$|{\mathbf n},{\mathbf n}\rangle$, we see that such terms obey a close equation~\cite{Dufour:PRL2012,Orso:PRL2005}: 
 \begin{equation}
 \label{integral}
\sum_{\mathbf m} K_{\mathbf n  \mathbf m} f_{\mathbf m} = \frac{1}{U}f_{\mathbf n},
 \end{equation} 
where  $K_{\mathbf n  \mathbf m}  =\langle {\mathbf n},{\mathbf n }|\hat G(E) |{\mathbf m},{\mathbf m}\rangle$. Hence, for a given energy $E$ of the pair, 
 Eq.(\ref{integral}) can be interpreted as an \emph{effective}  single-particle Schrodinger problem with eigen-energy $\lambda=1/U$.
The main purpose of this work  is to compute the associated mobility edge $U_c(W)$, for $E=0$.

We start by considering a 3D grid with transverse size $M$ and longitudinal size $L$. Differently from the 3D Anderson model, the matrix
$K$ of the effective Hamiltonian is dense and its elements have to be calculated numerically by expressing them in terms of the eigenbasis 
of the single-particle model, $\hat H^\textrm{sp} | \phi_r\rangle=\varepsilon_r  | \phi_r\rangle$:
\begin{equation}\label{KE}
K_{\mathbf n  \mathbf m} = \sum_{r=1}^N \phi_{\mathbf n r}  \phi_{\mathbf m r}^*  \langle \mathbf n | G^\textrm{sp}\mathbf (E-\varepsilon_r) | {\mathbf m}\rangle,
\end{equation}
where $G^\textrm{sp}(\varepsilon)=(\varepsilon I -H^\textrm{sp})^{-1}$ is the associated matrix resolvent, $I$ is the identity matrix and 
$\phi_{\mathbf n r} =\langle \mathbf n | \phi_r\rangle$. The eigenbasis is calculated by imposing open boundary conditions along the bar and
periodic boundary conditions in the transverse directions. 
 We see from Eq.(\ref{KE}) that the computation of the matrix $K$
 requires $N$ inversions of $N\times N$ matrices,  $N=M^2L$ being the total number of sites. 
 The matrix inversion is efficiently performed via recursive techniques~\cite{Jain:2007}, exploiting the
  block tridiagonal structure of the Hamiltonian (\ref{Anderson3D}). This allows to reduce the number of elementary operations
  from  $N^3$,  holding for a general matrix, to $M^6 L^2$. Hence the total cost for the evaluation of  $K$ 
  scales as $M^8 L^3$, which broadly exceeds  the cost $M^6 L$ of transfer matrix simulations for the same grid~\cite{McKinnonKramer:TransferMatrix:ZPB83}.
 This drastically limits the system sizes that we can explore. 
  In our numerics we keep the length of the bar fixed to $L=150$ and vary the transverse size $M$ between $8$ and $17$.

\section{NUMERICAL DETERMINATION OF THE CRITICAL POINT}
The  logarithm of the transmission amplitude of the pair, evaluated at a  position $n_z$ along the bar,  is given by~\cite{McKinnonKramer:TransferMatrix:ZPB83}:
\be\label{logT}
F(n_z)=\ln \sum_{\mathbf m_\perp,\mathbf n_\perp} |\langle \mathbf m_\perp,1| G^{\textrm p}(\lambda )| \mathbf n_\perp,n_z \rangle  |^2,
\ee
where $G^{\textrm p}(\lambda)=(\lambda I -K)^{-1}$ is the matrix resolvent of the effective model,
$\mathbf m_\perp =(m_x,m_y) $ and $\mathbf n_\perp= (n_x, n_y)$. In the limit $L\gg M$ the function (\ref{logT})
approaches a straight line, whose slope $p$ determines the Lyapunov exponent $\gamma$ according to $\gamma=-p/2$. 
The reduced localization length, needed for the finite-size scaling analysis,  is defined as $\Lambda_M=1/(\bar \gamma M)$, where  $\bar \gamma $ is the disorder-averaged Lyapunov exponent.

\begin{figure}
\includegraphics[width=\columnwidth]{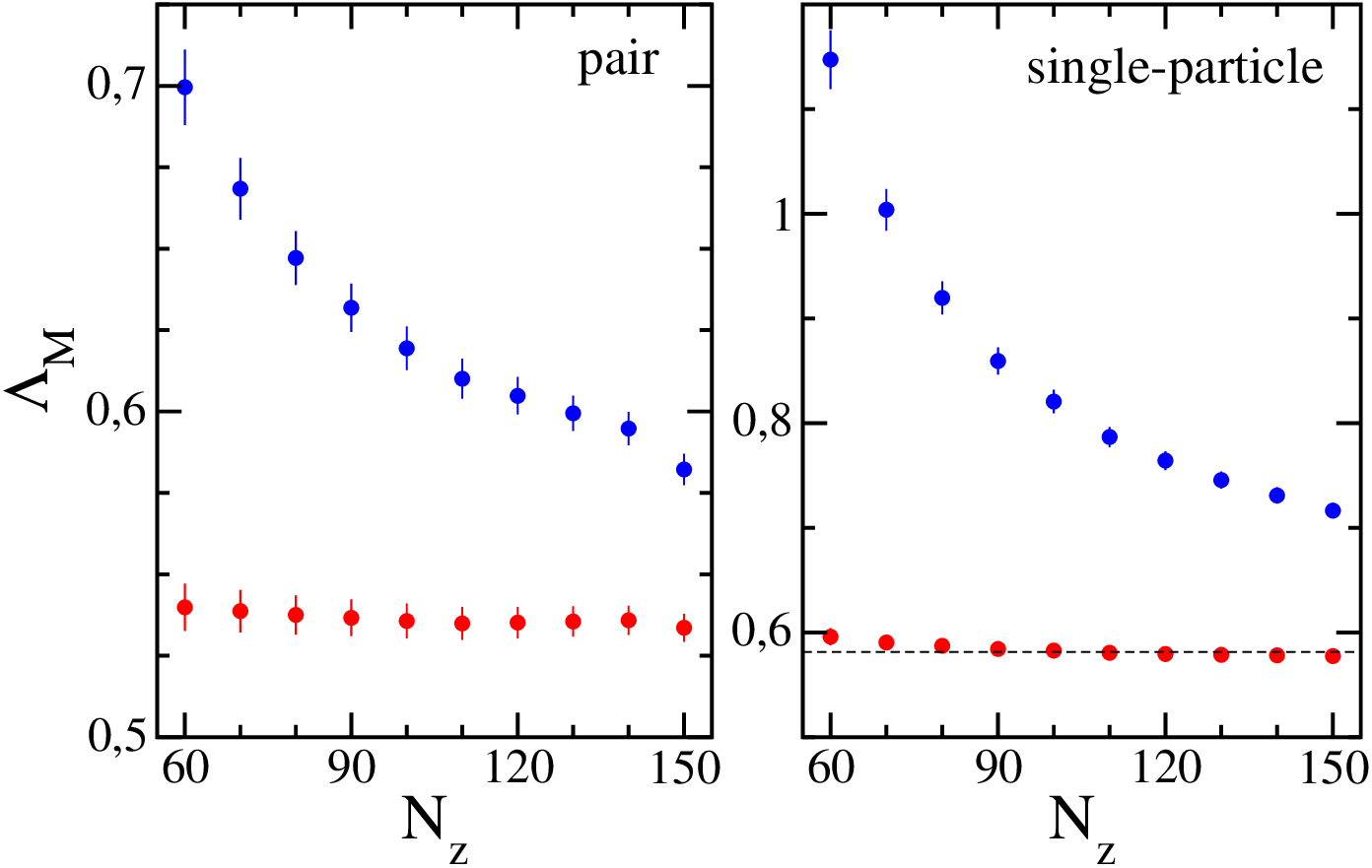}
\caption{ \emph {Left panel}: convergence study of the reduced  localization length $\Lambda_M=1/(\bar \gamma M)$ of the pair as a function of the position $N_z$ along the bar. 
Here $\bar \gamma$  denotes  the  Lyapunov exponent, averaged over $N_{tr}=701$  different disorder realizations, while the length and the transverse
size of the bar are $L=150$ and $M=12$, respectively. 
%\emph{Left panel}: Convergence study of two alternative methods to extract the reduced  localization length $\Lambda_M=1/(\bar \gamma M)$ of the pair 
%from a short bar of length $L=150$. 
%Here $\bar \gamma$  denotes  the  Lyapunov exponent averaged over $N_{tr}=701$ different disorder realizations and $M=12$ is the transverse size of the bar. 
The upper curve is obtained by calculating the Lyapunov exponent, for each disorder realization, via  $\gamma=-F(N_z)/(2 N_z)$, where $F$ is defined in Eq.(\ref{logT}). % and $N_z$ is the position along the bar. 
 The lower curve is instead obtained by fitting the data $(n_z,F(n_z))$ with $n_z=10, 20, ..., N_z$  by a straight line, $f_{fit}(n_z)=p n_z+q$,  and setting $\gamma=-p/2$. Only the fitting method yields converged results.
 The total energy of the pair is $E=0$, while the Hamiltonian parameters are  $W=23.5$ and $U=2$. 
 \emph{Right panel}: same  analysis but for the single-particle Anderson model, Eq.(\ref{Anderson3D}), for $\varepsilon=0$ (middle of the band) and $W=16.5$. The values of $L,M$ and $N_{tr}$ are 
the same as in the left panel. The dashed line corresponds to the estimate $\Lambda_M=0.5814\pm 0.0004$  obtained from transfer-matrix calculations performed 
on a bar of length $L=10^5$ after averaging over $240$  disorder realizations. }
\label{Fig:ConvergenceLambda}
\end{figure}
\begin{figure*}
\includegraphics[width=2\columnwidth]{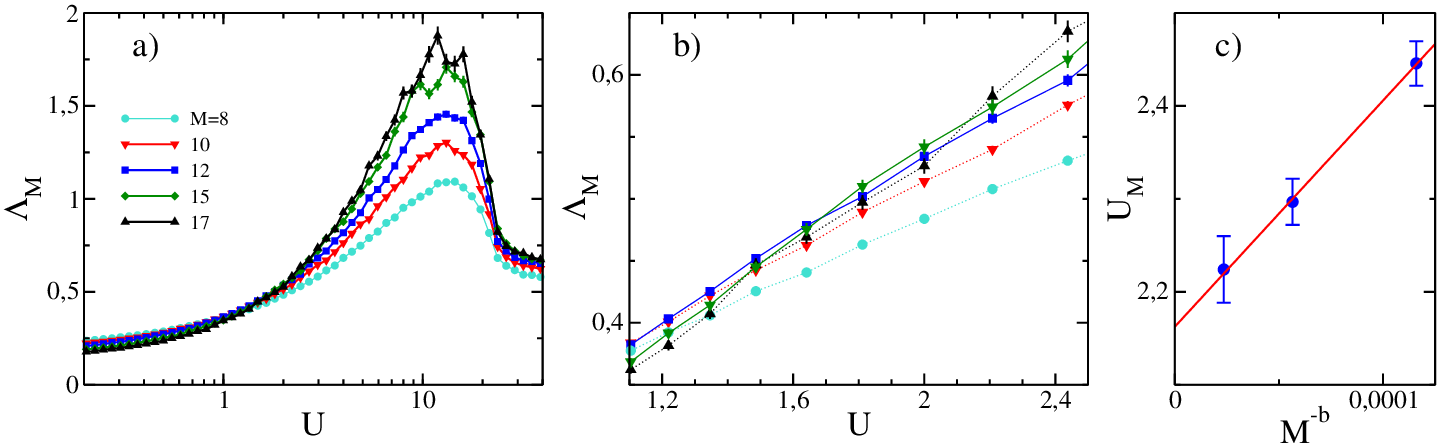}
\caption{a) %Numerical procedure to identify the critical point of the two-particle Anderson transition: the 
Reduced localization length as a function of 
the interaction strength for increasing values of the transverse size  $M=8, 10, 12, 15, 17$ of the bar, calculated using the fitting method. 
The energy of the pair is $E=0$ and the disorder strength is $W=23.5$, 
implying that all single-particle states are localized. The transition takes place at the point where all data curves with sufficiently large $M$ cross. 
%This point  drifts towards stronger interactions and upwards as $M$ increases, due to finite size effects. 
b) Zoom of the  region containing the crossing points for the largest system sizes. 
To improve visibility,  data points for $M=8, 10, 17$  are connected by dotted lines. 
%The critical exponent is consistent with the orthogonal universality class.
c) Numerical determination of the critical point: 
 the value $U_M$, defined by  $\Lambda_M(U=U_M)= \Lambda_{c}$, is plotted as a function of $M^{-b}$ for  $M=10,12,15$. For the orthogonal class $\Lambda_{c}=0.576$ and $b=3.94\pm 0.6$.  
 The straight line represents a  fit to the data, whose intercept yields the critical value $U_c=2.16\pm 0.04$. 
}
\label{Fig:crossing}
\end{figure*}
 In order to extrapolate it to  $L\rightarrow +\infty$ from our short bar, we proceed as follows. For each disorder realization, we evaluate 
$F(n_z)$ at regular intervals along the bar and extract the  slope by a linear fit,  $f_{fit}(n_z)=p n_z+q$. 
%, say $n_z=10k$, with $k=1,...,15$.  We then extract the slope by a linear fit,  $f_{fit}(n_z)=p n_z+q$. 
For a given position $N_z$ along the bar, we calculate  the slope by fitting only data points with $n_z\leq N_z$.
The  results calculated for $M=12, W=23.5$ and $U=2$ are displayed in the left panel of Fig.\ref{Fig:ConvergenceLambda} (bottom data curve). We see that 
the curve is rather flat as $N_z$ approaches $L$, suggesting that our fitting procedure is correct (see Supplemental Material~\cite{mynote}).
For comparison, in Fig.\ref{Fig:ConvergenceLambda} we also show (upper curve) the unconverged results 
%The upper curve in Fig.\ref{Fig:ConvergenceLambda} is instead 
obtained by using $p=F(N_z)/N_z$ (upper curve).

The right panel of  Fig.\ref{Fig:ConvergenceLambda} presents the same analysis for  the single-particle Hamiltonian (\ref{Anderson3D}) at zero energy, 
$\varepsilon=0$, and $W=16.5$. 
This is done by replacing $G^{\textrm p}$ with  $G^\textrm{sp}$ in Eq.(\ref{logT}), keeping unchanged the size of the bar as well as the number of disorder realizations.  
 The results based on the fitting method agree fairly well with the very accurate estimate obtained from transfer-matrix calculations (dashed line). 

The critical point of the metal-insulator transition can be identified by  studying the
behavior of $\Lambda_M$  as a function of the interaction strength $U$ and for increasing values of  the transverse size $M$.
 In the metallic phase, $\Lambda_M$ increases  with  $M$,  while in the insulating regime 
 $\Lambda_M$ decreases for $M$ large enough.  Exactly at the critical point $\Lambda_M$  becomes scale-invariant, that is 
 $\lim_{M\rightarrow +\infty} \Lambda_M=\Lambda_c$, where   $\Lambda_c$ is a constant of order unity, 
which only depends on the universality class of the model and on the specific choice of boundary conditions. For example 
the  Anderson model (\ref{Anderson3D}) belongs to the orthogonal universality class, 
where $\Lambda_{c,orth}= 0.576$ assuming periodic boundary conditions in the transverse directions.

In Fig.\ref{Fig:crossing} (panel a) we plot  our numerical results for the reduced localization length  
as a function of the  interaction strength  for increasing values of  $M$ assuming $W=23.5$, so that
 all single-particle states are localized.
Since  $E=0$,  the value of $\Lambda_M$ is independent of the sign of $U$, so hereafter we assume $U>0$.  
We see that interactions favor the delocalization of the pair and lead to an Anderson transition  around $U=2$.

Identifying the precise position of the critical point is not straightforward, because  the crossing point drifts towards stronger interactions and upwards as 
$M$ increases, due to finite size effects.
Simulating systems with even larger values of $M$ is computationally prohibitive: the data for $M=17$, obtained 
by averaging $N_{tr}=470$ disorder realizations, required already 700000 hours of computational time on a state-of-the-art supercomputer, 
and the curve is not smooth.

 As shown in the inset of Fig.(\ref{Fig:crossing}) (panel b), the height of the crossing point for the largest system sizes
(couples $M=12,17$ and $M=15,17$) becomes closer and closer to $\Lambda_{c,orth} $, suggesting that  also the effective model for the pair  belongs to the orthogonal 
universality class. In this case, no significant further drift  is expected.
 To verify this hypothesis, we need to compute the critical exponent $\nu$ related to the divergence of the localization length at the critical point, $\xi\sim|U-U_c|^{-\nu}$, and compare it with the numerical value $\nu_{orth} = 1.573$ known~\cite{Slevin:CriticalExponent:NJP14} for the orthogonal  class. 
 
According to the one parameter scaling theory of localization and for large enough $M$, the reduced localization length 
can be written in terms of a scaling function $f$ as
\begin{equation}\label{slevin1}
\Lambda_M =f (u(\omega) M^{1/\nu}),
\end{equation}
where $u$ is a function of the variable $\omega=(U-U_c)/U_c$, measuring the distance from the critical point. 
 Close to it, we can expand  the scaling functions $u$ and $f$ in Eq.(\ref{slevin1}) in Taylor series up to orders $m$ 
and $n$, respectively, as  $u(\omega)=\sum_{j=0}^{m} b_j \omega^j$ and $f(x)=\sum_{j=0}^{n} a_j x^j$. Following~\cite{Slevin:CriticalExponent:NJP14}, 
we set $b_1$=0, $a_1=0$ and $a_0=\Lambda_c$. The coefficients $a_j$ and $b_j$, as well as $U_c$ and $\nu$, are then obtained via 
a multilinear fit.
 We extract the critical exponent by fitting the (smoothest) data for $M=12$  and $M=15$  in the inset of Fig.\ref{Fig:crossing} with the ansatz (\ref{slevin1}). 
The latter should in principle include also irrelevant variables,  describing the drift of the crossing 
point. However, unlike $U_c$ and $\Lambda_c$, the value of the 
critical exponent is much less sensitive to these variables. 
 For $n=m=2$ we obtain 
 $\nu=1.64 \pm 0.13$, in full agreement with the universal value. All other crossings yield consistent results for $\nu$.
  
 Having found that on-site  interactions do not change the universality class of the transition,
 we can use this information to estimate $U_c$.
 Let  $U_M$ be the value of the interaction strength at which $\Lambda_M(U=U_M)=\Lambda_{c,orth}$.  
 For sufficiently large $M$, one can show~\cite{Campostrini:PRB2001} that 
  $U_M=U_c+a M^{-b}$, where $a$ is a numerical constant and $b=1/\nu_{orth}+y_{orth}$. Here $y_{orth}$ is  the leading irrelevant variable, whose value
 is also universal and given by $y_{orth}=3.3\pm 0.6$~\cite{Slevin:CriticalExponent:NJP14}. In Fig.(\ref{Fig:crossing}) (panel c) we show that the values of $U_M$
 extracted from our data curves for $M=10,12,15$ do vary linearly as a function of $M^{-b}$. 
 A linear fit to the data then yields $U_c=2.16\pm 0.04$.

\section{Phase diagram} 
 
\begin{figure}
\includegraphics[width=0.95\columnwidth]{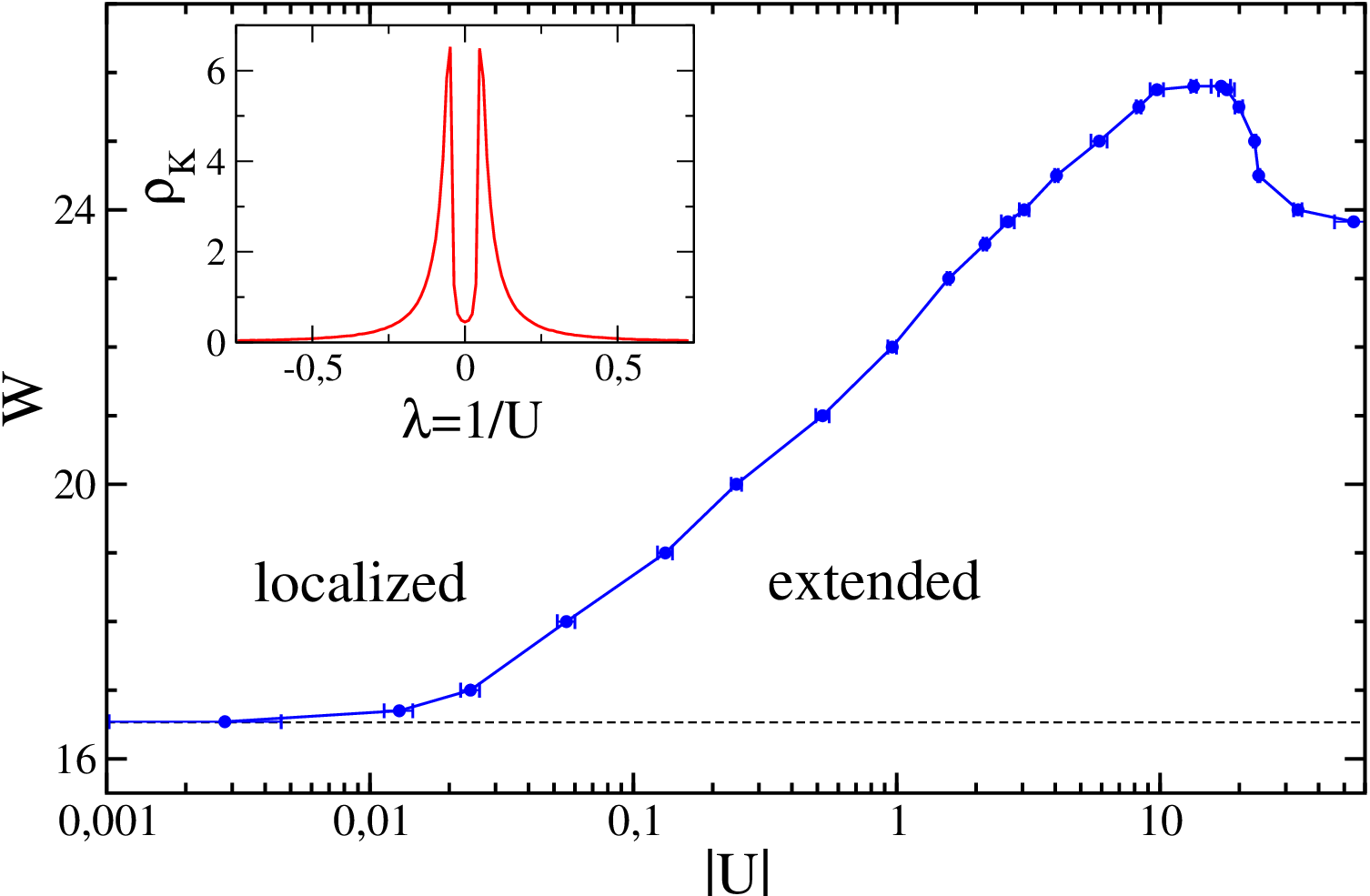}
\caption{\emph {Main panel:} phase boundary between localized and extended states in the $(U,W)$ plane, computed for a pair
with zero total energy, $E=0$. The dashed horizontal  line corresponds to the noninteracting limit, $W=W_c^{sp}=16.54$.
%where single-particle states in the middle of the band, $\varepsilon=0$,  become localized. %For $W<W_c^{sp}$ all eigenstates of the effective Hamiltonian are extended. 
 The diagram holds for both attractive and repulsive interactions. \emph {Inset:} disorder-averaged density of states $\rho_K$ of the effective Hamiltonian of the pair  
  calculated for $W=23.5$ using a cubic grid of sizes  $L=M=20$ with  periodic boundary conditions.}
\label{Fig:Wc}
\end{figure}

Next, we map out the  phase boundary  between localized and extended states of the pair in the $(U,W)$ plane.  
For each value of the disorder strength, we calculate the reduced localization length as a function of $U$ for $M=10,12,15$ 
and extrapolate the critical point from the scaling behavior of the $U_M$ values. To save computer resources,  we 
have limited the number of disorder realizations resulting in larger error bars for $U_c$.
 Moreover, for $W\leq 21$, we have  calculated the intercept by discarding also the data for $M=10$, as the relative deviation
$(U_M-U_c)/U_c$ increases as $W$ decreases.  

The  obtained results are displayed in Fig.\ref{Fig:Wc}. 
%For weak interactions, the critical disorder strength increases sharply from its noninteracting value, $W_c^{sp}$. 
We see that the Anderson transition for a pair with zero total energy occurs in a region, where all single-particle states are
 localized (see also Supplemental Material~\cite{mynote}). %In other words,  Moreover, the critical disorder strength increases sharply as a function of the interaction.
For $ 23.7\leq W \leq 25.9 $ the system possesses two distinct critical points,
resulting in a nonmonotonic behavior of the phase boundary. This  is best explained by calculating the disorder-averaged density of states of the effective model, 
$\rho_K(\lambda)=\overline{\sum_r \delta(\lambda-\lambda_r)/N}$,
%   \sum_{r=1}^N \sum_{i=1}^{N_{tr}} \delta(\lambda-\lambda_{r,i})/(N N_{tr})$,   
$\lambda_r$  being the eigenvalues of the kernel $K$. The result for $W=23.5$ is displayed in the inset 
of  Fig.\ref{Fig:Wc}. We see that $\rho_K$ is strongly peaked 
 at finite values of $\lambda$  and exhibits vanishing (power-law) tails. 
 This can be understood starting from  
 the strongly disordered limit,  $W \gg 1$. Since hopping terms can be neglected, the kernel $K$
becomes diagonal,  $K_{\mathbf n  \mathbf m}=\delta_{\mathbf n  \mathbf m}/(E-2V_\mathbf n)$, implying that
 \begin{equation} 
 \rho_K(\lambda)= \frac{1}{2W\lambda^2}\Theta\left(W-\left|E-\frac{1}{\lambda}\right|\right),
% \rho_K(\lambda)= \Theta(W-|E-1/\lambda|)/(2W\lambda^2),
 \end{equation} 
 where $\Theta$ is the  unit step function. 
 In particular for $E=0$ the density of states vanishes  for $|\lambda|<1/W$. Indeed,
 in order to interact, the two particles must lie on the same site $\mathbf n$, so the total energy is given by
 $E=U+2V_\mathbf n=0$, implying  $|U|= 2 |V_\mathbf n|\leq W$. 
 Reducing the disorder strength allows for tunneling between neighboring sites and leads to a finite value of $\rho_K(0)$, 
 as shown in the  inset of  Fig.\ref{Fig:Wc}.
 From the above discussion, one expects that weakly interacting states are the first to be  localized by disorder,
  whereas states with $|U|\sim W $  are the most robust against  localization, in agreement with the phase diagram of Fig.\ref{Fig:Wc}. 
  
It is worth mentioning that a nonmonotonic behavior of the critical disorder strength versus $U$ was also obtained 
for the \emph{ground state}  of the Anderson-Hubbard model at finite fillings in earlier 
theoretical studies~\cite{Byczuk:PRL2005}~\cite{Henseler:PRB2008} based, respectively, on the dynamical mean field theory and 
on the self-consistent theory of localization.  

%Finally, for $W<W_c^{sp}$, tunneling effects are so strong that \emph{all} states of the effective model are extended  
 %(see Supplemental Material~\cite{mynote}). 

%We see from Fig.\ref{Fig:Wc} that the critical disorder strength increases sharply from its noninteracting value, $W_c^{sp}$.  

%Finally, as shown in Fig.\ref{Fig:Wc},  for $W<W_c^{sp}$ \emph{all} states with zero total energy are extended  (see Supplemental Material~\cite{mynote}).
%This is a specific feature of the effective Hamiltonian of the pair, not shared by the single-particle Anderson model.   
 
 While interactions favor the delocalization of  pair states with $E=0$, their effect on tightly bound states, corresponding to $E\simeq U \rightarrow \infty$,
 is the opposite. 
 % the situation is \textsl{reversed} for tightly bound states, corresponding to $E\simeq U \rightarrow \infty$. 
 As discussed in~\cite{Dufour:PRL2012}, these states obey the single-particle model (\ref{Anderson3D})
with renormalized  disorder strength $W_m=2W$ and strongly reduced tunneling rate $J_m=2 J^2/|U|$, implying that they are localized by a weak disorder,   
$W_c=16.54J^2/|U|$.  
 
 \section{CONCLUSIONS AND PERSPECTIVES}
 To summarize, we have studied the localization properties of  two interacting particles in the 3D Anderson-Hubbard model. 
 Based on large scale numerical calculations,  we have computed the phase boundary separating 
 localized from extended states in the $(U,W)$ plane for zero total energy of the pair. 
 We have shown that the effective two-body mobility edge lies in a region where all single-particle states are localized. In particular the critical 
 disorder strength depends nonmonotonically on $U$ and features a sharp enhancement for weak interactions. We interpret this result from the behavior 
 of the disorder-averaged density of states of the effective model.

%It would be interesting to apply our approach to  attractively $(E<2\varepsilon_1)$ and repulsively $(E>2\varepsilon_N)$ bound states, for which Anderson localization sets in already for  modest values of the disorder strength, as  previously found~\cite{Dufour:PRL2012} in one-dimensional quasi-periodic lattices.
 % Hence, we foresee that the complete phase diagram in the $(E,U,W)$ space will show a very rich structure. 
Our theoretical results can be addressed in current experiments with ultra-cold atoms~\cite{Krinner:PRL2015}. They also provide  a solid test-bed for future studies of
mobility edges in 3D many-body systems.
Finally, our numerical method can also be  adapted to investigate  the localization of Cooper pairs in strongly disordered  
superconductors~\cite{Feigelman:PRL2007,Sacepe:NatPhys2011}.

\section*{ACKNOWLEDGEMENTS} 
   We acknowledge fruitful discussions with D. Basko.
This project has received funding from the European Union's Horizon 2020 research and innovation programme under the 
Marie Sklodowska-Curie grant agreement No 665850. This work was granted access to the HPC resources of CINES (Centre Informatique National de l'Enseignement Sup\' erieur) under the allocations 
2016-c2016057629, 2017-A0020507629 and 2018-A0040507629 supplied by GENCI (Grand Equipement National de Calcul Intensif).

\bibliographystyle{apsrev}
\bibliography{ArtDataBasev4}

%\clearpage
%\includepdf{supplemental_2body.pdf}

\end{document}

\title{Supplemental material for "Mobility edge of two interacting particles in three-dimensional random potentials"}

\maketitle 

%\begin{minipage}{0.45\linewidth}

%\vspace{0.5cm}
 \textbf{I. Fitting procedure}  
 \vspace{0.15cm}

In the left panel of Fig. 1 of the main text, we have shown that the fitting method yields converged results for the reduced localization length 
for a specific value $U=2$ of the interaction strength, close to the critical point, $U_c\simeq 2.16$. 
For completeness, in the Fig.1 below  we display $\Lambda_M$ versus $U$ for increasing values of the position  $N_z$ along the bar. We see that effects due to the shortness 
of the bar are small and appear well inside the metallic region, where they lead to overestimate the correct result.
 \vspace{0.35cm}
 \begin{center}
 %\begin{figure}
 \includegraphics[width=0.925\columnwidth]{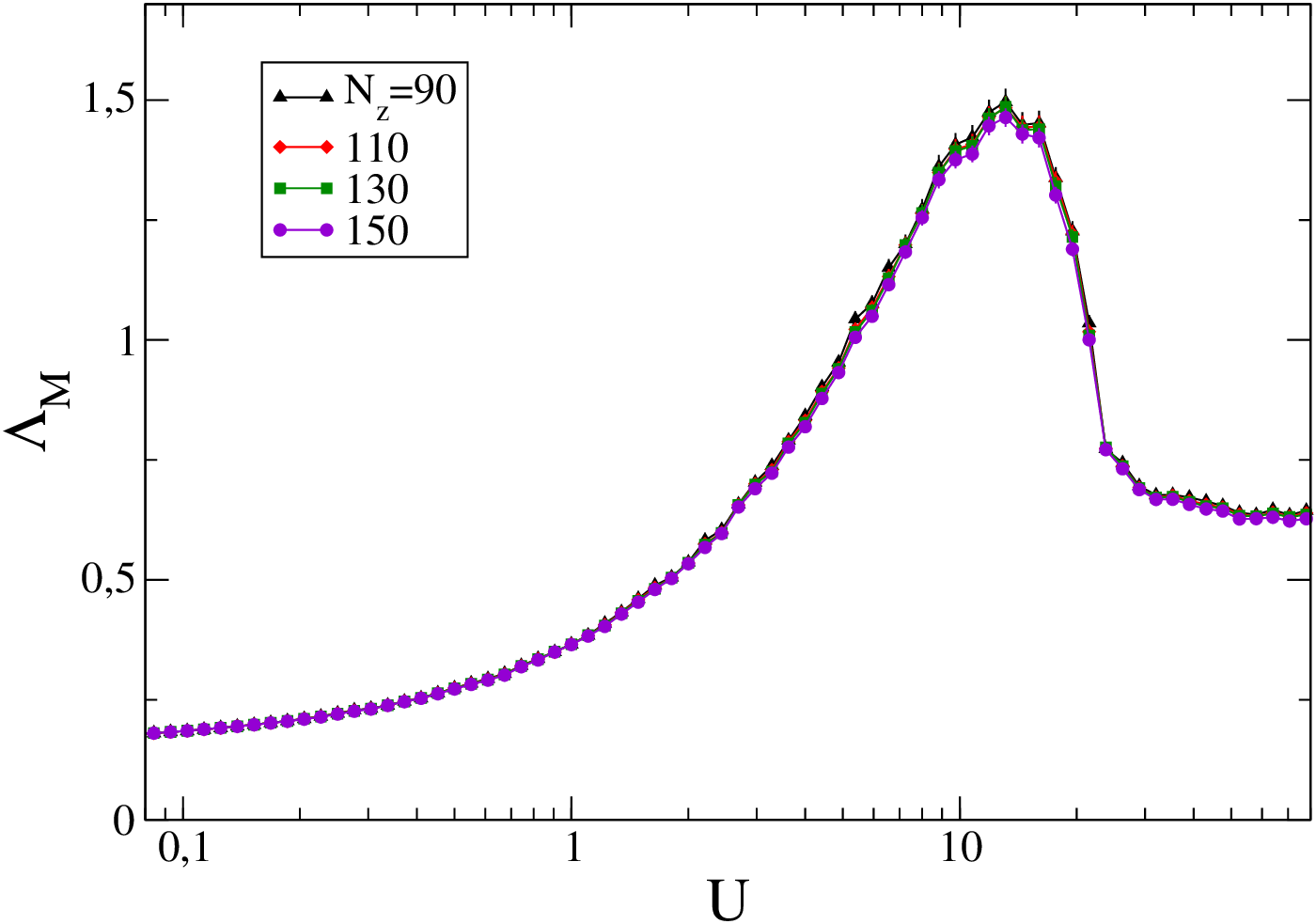}
%  \caption{\small{Reduced localization length of the pair, calculated via the fitting method,  as a function of the interaction strength $U$ for increasing values $N_z=90, 110, 130, 150$ of the position $N_z$ along the bar.  The other parameters are the same as in the left panel of Fig.1 in the main text.}}
%\label{fig:sm0}
%\end{figure}
 \end{center}
\begin{minipage}{\columnwidth}
\small{FIG.1 Reduced localization length of the pair, calculated via the fitting method,  as a function of the interaction strength $U$ for increasing values $N_z=90, 110, 130, 150$ of the position $N_z$ along the bar.  The other parameters are the same as in the left panel of Fig. 1 in the main text.}
\end{minipage}
 
\vspace{0.75cm}
 \textbf{II. Phase diagram}  
\vspace{0.12cm}

We provide the reader with some additional information concerning  the phase diagram presented in Fig. 3 of the main text.
For disorder strengths such that $ 23.7\leq W \leq 25.9 $, the system possesses two distinct critical points. As an example, in Fig. 2  we 
display $\Lambda_M$ versus $U$ calculated at $W=24.5$ for three different values  $M=10, 12, 15$
 of the  transverse size of the bar. We indeed distinguish two critical points, $U_c=4.04$ and $U_c=23.7$. The system 
  possesses delocalized states with zero total energy only for interaction strengths between these two values. 
As $W$ increases, the two critical points get closer and closer, until they merge around $U=14$ for $W=25.9$.

\vspace{0.5cm}

%\begin{wrapfigure}{c}{1.0\columnwidth}
%\end{wrapfigure}

%\begin{wrapfigure}{c}{1.0\columnwidth}
  \begin{center}
 % \begin{figure}
    \includegraphics[width=0.925\columnwidth]{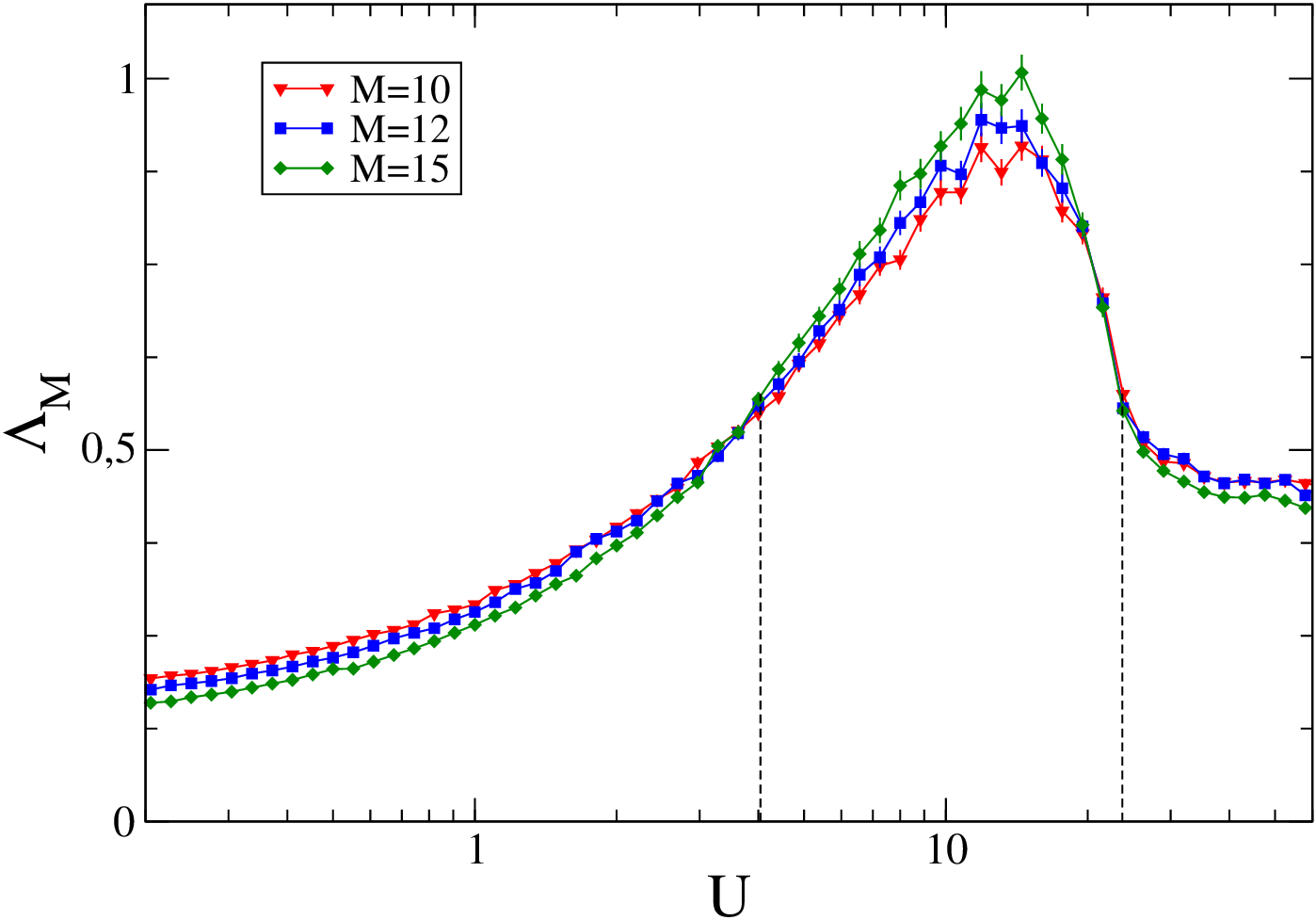}
\label{fig:sm1}
%\end{figure}
 \end{center}
%\end{wrapfigure}
\begin{minipage}{\columnwidth}
 \small{FIG. 2 Reduced localization length of the pair as a function of the interaction strength $U$ calculated at $W=24.5$ and 
 total energy $E=0$ for three different values of the transverse size
of the bar, $M=10, 12, 15$. The dashed lines mark the position  of the two critical points, $U_c=4.04$ and $U_c=23.7$. The length of the bar is $L=150$ and the number of  disorder realizations is $N_{tr}=200$.}
\end{minipage}

\vspace{0.6cm}

For $W<16.54$, all pair states with zero total energy are extended, as illustrated  in Fig. 3 for $W=14$. Indeed 
$\Lambda_M$ grows with system size  for any value of the interaction strength, implying diffusive behavior.

\vspace{0.3cm}

%\begin{wrapfigure}{c}{1.0\columnwidth}
  \begin{center}
 %  \begin{figure}
    \includegraphics[width=0.925\columnwidth]{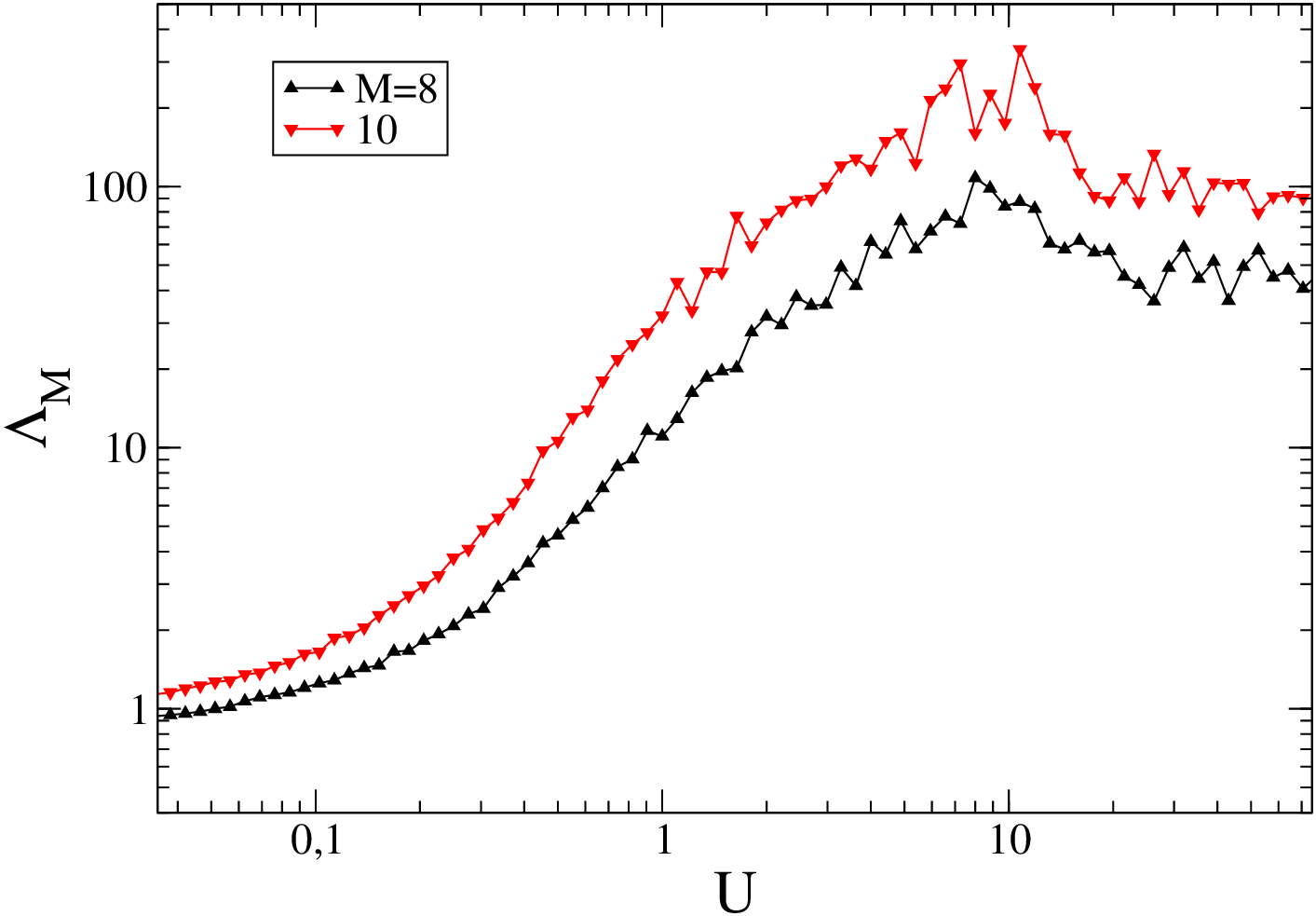}
 \label{fig:sm2}
%\end{figure}
 \end{center}
%\end{wrapfigure}
\begin{minipage}{\columnwidth}
\small{FIG. 3 Reduced localization length as a function of the interaction strength $U$ calculated at $W=14$ and 
 total energy $E=0$ for two different values of the transverse size
of the bar, $M=10$ and $M=12$. The results are obtained by averaging over $N_{tr}=100$ different realizations of the random potential.
 The length of the bar is $L=150$.}
 \end{minipage}

%\vspace{0.7cm}

%\begin{figure}
%\includegraphics[width=0.9\columnwidth]{fig1Sup.eps}
%\caption{
%     Reduced localization length of the paper as a function of the interaction strength $U$ calculated for $W=24.5$ and for two different values of the transverse size
%	of the bar, $M=12$ and $M=15$. The two curves cross at two distinct (critical) points $U_c=4.2$ and $U_c=22.6$. The length of the bar is $L=150$ and the number of 
%	disorder realization is $N_{tr}=200$. }
%	\label{fig1:sm}
%\end{figure}
%
%%\captionsetup{width=0.8\columnwidth}
%%\centering
%
%\begin{figure}
%	\includegraphics[width=0.9\columnwidth]{fig2inset.eps}
%	\caption{
%	Reduced localization length of the paper as a function of the interaction strength $U$ calculated for $W=14$ and for two different values of the transverse size
%	of the bar, $M=10$ and $M=12$. The system is diffusive for all values of the ehibits metallic two curves never cross and all states of the effective model are metallic. The length of the bar is $L=150$ and the number of 	disorder realization is $N_{tr}=100$.}
%	%Notice that there are 6 samplings points between the origin and the first zero of the correlation function, which occurs at $r/\sigma=3.83171$. 
%	%This is enough to faithfully reproduce the correlation function of the continuous model.}
%	\label{fig2:sm2}
%\end{figure}

\bibliographystyle{apsrev}
\bibliography{ArtDataBasev4}